Concept: Complexity
Title: Unifying 'almost everything'
Box: The laws describing the behaviour of a complex system are qualitatively different from those describing its units.

Author: Tamas Vicsek

If a concept is not well defined, there are grounds for its abuse. This is particularly true of complexity, an inherently interdisciplinary concept that has penetrated very different fields of intellectual activity from physics to linguistics, but with no underlying, unified theory. Complexity has become a popular buzzword used in the hope of gaining attention or funding -- institutes and research networks associated with complex systems grow like mushrooms. Why and how did it happen that this vague notion has become a central motif in modern science? Is it only a fashion, a kind of sociological phenomenon, or is it a sign of a changing paradigm of our perception of the laws of nature and of the approaches required to understand them? Because virtually every real system is inherently extremely complicated, to say that a system is complex is almost an empty statement – couldn't an Institute of Complex Systems just as well be called an Institute for Almost Everything? Despite these valid concerns, the world is indeed made of many highly interconnected parts over many scales, whose interactions result in a complex behaviour needing separate interpretation for each level. This realization forces us to appreciate that new features emerge as one goes from one scale to another, so it follows that the science of complexity is about revealing the principles governing the ways by which these new properties appear.

In the past, mankind has learned to understand reality through simplification and analysis. Some important simple systems are successful idealizations or primitive models of particular real situations, for example, a perfect sphere rolling down on an absolutely smooth slope in vacuum. This is the world of newtonian mechanics, and involves ignoring a huge number of simultaneously acting other factors. Although it might sometimes not matter if details such as the billions of atoms dancing inside the sphere's material are ignored, in other cases reductionism may lead to incorrect conclusions. In complex systems, we accept that processes occurring simultaneously on different scales or levels matter, and the intricate behaviour of the whole system depends on its units in a non-trivial way. Here, the description of the behaviour of the whole system requires a qualitatively new theory, because the laws describing its behaviour are qualitatively different from those describing its units.

Take, for example, turbulent flows and the brain. Clearly these are very different systems, but they share a few remarkable features, including the impossibility of predicting the rich behaviour of the whole by merely extrapolating from the behaviour of its units. Who can tell from studying a little drop or a single neuron what are the laws describing the intricate flow patterns in turbulence or the electrical activity patterns produced by the brain? Moreover, both these systems (and many others) are such that randomness and determinism are both relevant to their global behaviour. They exist on the edge of chaos; they are able to produce nearly regular behaviour, but also can change dramatically and stochastically in time and/or space as a result of small changes in conditions. This seems to be a general property of systems capable of producing interesting (complex) behaviour.

Knowledge of the physics of elementary particles is therefore useless for these higher scales. Entering a new level or scale is accompanied by new, emerging laws governing it. When creating life, nature acknowledged the existence of these levels by spontaneously separating them as molecules, macromolecules, cells, organisms, species and societies. The big question is whether there is a unified theory for the ways elements of a system organize themselves to produce a behaviour typical for wide classes of systems. Interesting principles have been proposed, including self-organization, simultaneous existence of many degrees of freedom, self-adaptation, rugged energy landscape and scaling (for example power-law dependence) of the parameters and the underlying network of connections. Physicists are learning how to build relatively simple models producing complicated behaviour, while those working on inherently very complex systems (biologists or economists, say) are uncovering the ways their infinitely complicated subjects can be interpreted in terms of interacting, well-defined units (such as proteins).

What we are witnessing in this context is a change of paradigm in attempts to understand our world as we realize that the laws of the whole cannot be deduced by digging deeper into the details. In a way, this change has been invoked by development of instruments. Traditionally, improved microscopes or bigger telescopes are built to understand better particular problems. But computers have allowed new ways of learning. By directly modelling a system made of many units, one can observe, manipulate and understand the behaviour of the whole system much better than before, as in networks of model neurons and virtual auctions by intelligent agents, for example. In this sense, a computer is a tool improving not our sight (as in the microscope or telescope), but our insight into mechanisms within a complex system. Further, use of computers to store, generate and analyse huge databases -- the fingerprints of systems that people otherwise could not comprehend.

Many scientists implicitly assume that the understanding of a particular phenomenon is achieved if a (computer) model provides results in good agreement with the observations and leads to correct predictions. Yet such models allow us to simulate systems far more complex than those that have solvable equations. In the newtonian world, exact or accurate solutions of the equations of motion provide predictions for future events. As a rule, models of complex systems result in a probabilistic prediction of behaviour, and the form in which conclusions are made is less rigorous compared to classical quantitative theories, even involving elements of 'poetry'.

Suggested reading:

M. M. Waldrop, "Complexity" (Simon and Schuster, New York, 1993)

M. Gell-Mann, "Plectics: the study of simplicity and complexity",
Europhysics News, 33/1 (2002) p. 17

http://www.comdig.org